\documentstyle[prb,aps,psfig]{revtex}

\thicklines

\begin{document}

\title{Noncollinear magnetic ordering in small Chromium Clusters}
\author{ C. Kohl, G.F. Bertsch}
\address{Institute for Nuclear Theory -- Department of Physics,\\
         University of Washington Seattle, WA 98195, USA}
\maketitle

\begin{abstract}
We investigate noncollinear effects in antiferromagnetically coupled clusters
using the general, rotationally invariant form of local spin-density theory. 
The coupling to the electronic degrees of freedom is treated with relativistic 
non-local pseudopotentials and the ionic structure is optimized by Monte-Carlo
techniques. We find that small chromium clusters (N $\le$ 13) strongly favor
noncollinear configurations of their local magnetic moments due to 
frustration. This effect is associated with a significantly lower total 
magnetization of the noncollinear ground states, ameliorating the 
disagreement between Stern-Gerlach measurements and previous collinear 
calculations for Cr$_{12}$ and Cr$_{13}$. Our results further suggest that 
the trend to noncollinear configurations might be a feature common to most 
antiferromagnetic clusters.\\

\vspace{-2mm}
\end{abstract}

\section{Introduction}
\label{intro}
\noindent
Magnetic properties of transition metal clusters have become the
subject of intensive research, both from the 
experimental~\cite{exp1,exp2,exp3,exp4} and theoretical point of 
view~\cite{theo1,theo2,theo3,theo4}. One of the most interesting
and challenging aspects of that field is the subtle interplay between 
geometric structure and magnetic ordering which has mostly been investigated
for ferromagnetic 3d-clusters and 4d-clusters. Finite size effects and a 
reduced dimensionality often lead to a significantly different magnetic 
behavior from the bulk with clusters having enhanced atomic 
moments\cite{rhodium,nickel}, larger anisotropy energies\cite{aniso} or an 
altered temperature dependence of the magnetization\cite{temp}.\\
Almost all theory to date employed the local spin-density approximation (LSDA)
with the assumption that the spin density matrix is diagonal in 
some particular frame. In that special case the spins are automatically 
collinear along a fixed quantization axis. The only generalized spin-density 
calculation for clusters that treats the electron spin as a vector observable 
and a function of position has recently been performed by Car and 
coworkers.\cite{car}. They have shown that noncollinear configurations exist 
in Fe$_{\rm 3}$ and Fe$_{\rm 5}$ although the effect on structure and 
energetics of these ferromagnetic (FE) clusters is not very pronounced. On the
other hand an unconstrained orientation of the quantization axis is known to 
play a key role in describing various non-ferromagnetic systems like the 
$\gamma$-phase of bulk iron\cite{bulkfe}, disordered systems\cite{disord} or 
ultra thin Fe films with a partial antiferromagnetic (AF) 
coupling\cite{nonfer}. Furthermore, the work on Fe/Cr (ref.\cite{multi1}) and 
Ag/Cr multilayers\cite{multi2} demonstrated how the competition between AF 
ordering and frustration of the Cr moments leads to noncollinear arrangements 
in form of a spin-density wave. Although the importance of frustration in AF 
systems seems to be evident and was discussed in detail for embedded Cr 
clusters by Pastor et al.\cite{embedd} and more generally by Manninen and 
coworkers\cite{mann}, the possibility of noncollinear effects has not yet been
considered for AF clusters.

In this paper, we present the first general local spin-density calculation 
for clusters of AF materials. Besides the motivation given above, we have 
decided to explore noncollinear effects in chromium clusters for two reasons: 
First, chromium is particularly challenging amongst the 3d-elements due to its
maximal spin alignment in the atom $(3d^54s^1)$: all valence electrons have 
parallel spins, i.e. the total magnetization is S=3. This leads 
to a variety of unique effects including an unusually short dimer bonding 
length or a repulsion for FE coupling at small distances\cite{dimerexp,dimer}.
The other reason why we chose chromium is the possibility to compare with the 
thorough {\it ab-initio} study of Cr clusters by Cheng and Wang\cite{cheng}.
These authors recently employed the conventional LSDA for N$_{\rm at} \le$ 15,
finding all clusters to be antiferromagnetically coupled. Their structures 
can serve as a benchmark to check our computations concerning the restriction 
to collinear configurations -- an aspect that is crucial to extract the 
importance of noncollinear effects.\\
In Section~\ref{theo} we outline the fundamentals of our theoretical 
approach and briefly motivate the structure optimization. This involves the 
rotationally invariant LSDA to minimize the electrons and a relativistic, 
non-local pseudopotential for their interaction with the ions. 
In Section~\ref{resdis} we discuss some numerical tests and present our 
results. We find noncollinear spin-configurations for all investigated 
clusters. We demonstrate the influence of noncollinearity on the geometry and 
on the total magnetic moment and discuss how this reduces the discrepancy
concerning the magnetization of some chromium clusters between the experiment 
by Bloomfield et al.\cite{bloom} and some previous theoretical 
results\cite{cheng,muprev}.

\section{Theoretical and Numerical Background}
\label{theo}
\noindent
The density functional theory in its most general form, as developed by 
Barth and Hedin\cite{hedin}, allows the orientation of each spin to vary with 
position. The wavefunctions are described by complex two-component 
spinors $\Psi = \left( \Psi_\sigma, \Psi_{\sigma^\prime} \right)$, where 
$\sigma$ and $\sigma^\prime$ denote the spin indices, and the degrees of 
freedom are the elements of the single-particle spin-density matrix 
\begin{equation}
  \rho_{\sigma \sigma^\prime} \left(\vec{r}\right) =
   \sum\limits_i \Psi_{i,\sigma} \left(\vec{r}\right) \,
                 \Psi^\ast_{i,\sigma^\prime}  \left(\vec{r}\right) \quad .    
\end{equation}
Assuming this matrix to be diagonal, the usual local spin-density 
functionals are parameterized in terms of $\rho_{\uparrow \uparrow}(\vec{r})
=:\rho_{\uparrow}(\vec{r})$ and $\rho_{\downarrow \downarrow}(\vec{r})
=:\rho_{\downarrow}(\vec{r})$ only. 
In that special case, the spins are necessarily collinear along the chosen
quantization axis and the exchange-correlation potential is obtained via
$V^{\rm xc}_{\sigma}=\partial E^{\rm xc}\left[\rho_{\uparrow},
\rho_{\downarrow}\right] / \partial \rho_{\sigma}$. However, rotational 
invariance requires that the true variables are the eigenvalues 
$n_\uparrow(\vec{r})$ and $n_\downarrow(\vec{r})$ of the spin density 
matrix $\rho_{\sigma \sigma^\prime} (\vec{r})$. We can thus apply  
standard local spin-density functionals (we chose the formula of Perdew and 
Wang\cite{perdew}) by evaluating the potential in a locally diagonal frame.
The transformation is carried out following the work of K\"ubler et 
al.\cite{kueb} who used the spin-1/2 rotation matrix 
\begin{equation}
\hat{U}\left(\vec{r}\right) = 
  \left( \begin{array}{cc}
\mbox{cos}\,\frac{\theta (\vec{r})}{2} \,e^{\frac{i}{2} \phi(\vec{r})} & 
\quad
\mbox{sin}\,\frac{\theta (\vec{r})}{2} \,e^{\frac{-i}{2} \phi(\vec{r})} \\
-\mbox{sin}\,\frac{\theta (\vec{r})}{2} \,e^{\frac{i}{2} \phi(\vec{r})} &
\quad
\mbox{cos}\,\frac{\theta (\vec{r})}{2} \, e^{\frac{-i}{2} \phi(\vec{r})}
   \end{array} \right) 
\end{equation}
to locally diagonalize the spin-density matrix: 
\begin{equation}
\label{diagmat}
\sum_{\sigma \sigma^\prime} U_{\alpha \sigma} (\vec{r}) \, \rho_{\sigma 
\sigma^\prime} (\vec{r}) \, U^\ast_{\sigma^\prime \beta}(\vec{r}) = 
\delta_{\alpha \beta}\, n_\alpha(\vec{r}) 
\end{equation}
By working in this representation we express $\partial E^{\rm xc}/ \partial 
\rho_{\sigma \sigma^\prime}$ by $E^{\rm xc}/ \partial n_{\sigma}$ plus the 
introduction of local spin rotation angles $\phi\left(\vec{r}\right)$ and 
$\theta \left(\vec{r}\right)$ which are the local azimuthal and polar
angles of the magnetization density vector. They are computed from 
eqn.(\ref{diagmat}) through the requirement of vanishing off-diagonal elements 
as
\begin{equation}
\label{theta}
\phi\left(\vec{r}\right) = -\, \mbox{tan}^{-1} \, \frac{\mbox{Im}\, 
\rho_{\uparrow \downarrow}\left(\vec{r}\right)}{\mbox{Re}\,  
\rho_{\uparrow \downarrow}\left(\vec{r}\right)} \,\, , \qquad \qquad
\theta\left(\vec{r}\right)  =  \mbox{tan}^{-1}\, \frac{2 \left[
\left( \mbox{Re}\rho_{\uparrow \downarrow}\left(\vec{r}\right) \right)^2 +
\left( \mbox{Im}\,\rho_{\uparrow \downarrow}\left(\vec{r}\right) \right)^2 
\right]^{1/2} } {\rho_{\uparrow \uparrow} \left(\vec{r}\right) - 
\rho_{\downarrow \downarrow} \left(\vec{r}\right)} \quad .
\end{equation} 
These new degrees of freedom complicate the mean-field equations and lead to
an exchange-correlation potential $\hat{V}^{\rm xc}$ in form of a complex 
matrix in spin-space
\begin{equation}
\hat{V}^{\rm xc} = \frac{1}{2} \left( V^{\rm xc}_{\uparrow} +
V^{\rm xc}_{\downarrow} \right) \hat{1} + \frac{1}{2}
\left( V^{\rm xc}_{\uparrow} - V^{\rm xc}_{\downarrow} \right) 
\hat{\vec{\sigma}} \cdot \vec{d} \quad ,
\end{equation}
where $\vec{d}$ is a position dependent unit vector along the direction
of the vector $\left(\mbox{Re}\, \rho_{\uparrow \downarrow}(\vec{r}), \,
\mbox{Im} \, \rho_{\uparrow \downarrow}(\vec{r}), \, 
\rho_{\uparrow \uparrow}(\vec{r}) - \rho_{\downarrow \downarrow}(\vec{r})
\right)$. The presence of the second term in the exchange-correlation
potential allows a general coupling of the up- and down-components of the 
spinor wavefunctions. The wavefunctions belong to the class TSDW of the 
Fukutome classification scheme\cite{fukutome}. To interpret the magnetic 
properties, we compute the vector magnetization density $\vec{m}(\vec{r})$ by 
expressing the spin-density matrix in the form
\begin{equation}
\hat{\rho}\left(\vec{r}\right) = 0.5 \left[ n(\vec{r}) \, 
 \hat{1} +
\vec{m}(\vec{r}) \cdot \hat{\vec{\sigma}} \right] \quad.
\end{equation}
We associate magnetic moments with individual atoms by integrating
each component of $\vec{m}(\vec{r})$ within a sphere centered on the ions,
giving us the local magnetic moment vectors $\vec{\mu}_{\rm at}$. The 
integration radius is chosen to be one half of the smallest interatomic 
distance in each cluster to avoid overlap and the resulting spheres contain 
about $80-90 \, \%$ of the magnetization density. Taking a larger radius for 
more distant atoms has a negligible effect on the orientation of their 
magnetic moment, although its magnitude becomes somewhat bigger. \\
As usual, we only treat the valence electrons explicitly, taking care of the 
ionic core with a pseudopotential approximation. We use the relativistic 
pseudopotential from Goedecker et al.\cite{goed}, which contains a local part
plus a sum of separable gaussians, optimized to represent a transferable 
non-local pseudopotential on a coordinate mesh. The multiplication of the 
wavefunctions with the non-local part can be limited to a small region around 
the ions as the radial projectors fall off rather quickly. However, 20 
integrations within the covalent radius of each atom need to be performed to 
correctly account for non-local effects in chromium. The energetics at small 
ionic separations inside the clusters further requires us to include the 3s 
and 3p semi-core electrons into the variational space.
Our pseudopotential also includes spin-orbit terms which fix the 
orientation of the total magnetization M to the ionic structure thus
giving rise to magnetic anisotropy. The implementation of the $\vec{L} \cdot 
\vec{S}$ operator is not too costly because we have to deal with a complex 
spinor-structure anyway. Spin-orbit effects enable us to test 
the validity of the usually applied atomic sphere approximation by studying 
the intra-atomic dispersion\cite{asa}. 

We have carried out an unconstrained structural search by fully minimizing 
electronic and ionic degrees of freedom. To find the ground state and stable 
isomers, the ionic positions were optimized by Monte-Carlo sampling applying 
the technique of simulated annealing. After some Metropolis steps, the 
electronic wavefunctions are updated with Kohn-Sham iterations.
The optimization of the ionic geometry involves a minimization of the one-ion 
energies and is explained in detail in ref.\cite{diss}.
The static Kohn-Sham equations are solved in a combined coordinate and 
momentum space representation by using an efficient damped gradient 
iteration\cite{damp}. Local operators are applied on coordinate space 
wavefunctions while the kinetic energy and the action of the spin-orbit 
operator are computed in momentum space applying fast Fourier techniques. 
The Poisson equation is solved via the FALR-method\cite{falr}. As it is 
more convenient for most physical observables, electronic wavefunctions 
and densities are stored on a three-dimensional coordinate space
mesh. We perform our calculations in a cubic box with a mesh spacing of 
0.32 a.u. and up to 64 grid points in each direction. We checked that the mesh
size was big enough to avoid artifacts from the boundaries. A detailed 
description of our numerics can be found in refs.\cite{numasp}.

\section{Results and Discussion}
\label{resdis}

\noindent
Before discussing our results we mention some of the various tests we
performed in order to increase our confidence in the Hamiltonian and its
numerical implementation.
The dimer plays a key role in the description of small chromium clusters.
It is known that its subtle electronic properties demand a high accuracy of 
the Cr-Cr interactions and the numerical representation\cite{dimerexp,dimer}. 
By applying the pseudopotential in the semi-core version, our binding energy 
(1.98 eV) and bonding length (d$_0= 3.25$ a.u.) for the antiferromagnetic 
ground state were in good agreement with experimental 
results\cite{dimerexp} (1.56 eV $\pm$ 0.3 eV, d$_{\rm exp} = 3.19$ a.u.) and 
previous all-electron or pseudopotential calculations\cite{dimer}. The correct 
energetic order of the single particle levels as a function of the 
intra-molecular distance, the symmetry of the wavefunctions and the properties 
of the ferromagnetic coupling (d $= 5.2$ a.u.) could be reproduced as well. 
The same holds for the bonding length of the CrO molecule which deviated from 
the experimental result by $1.3 \, \%$. Additionally, we achieved degenerate 
d-states up to a level of $1 \, \%$ and the correct energetic order of 3d and 
4s-levels in the chromium atom. The rotationally invariant 
spin-density theory was checked by letting the FE configuration of the dimer 
relax to the AF ground state. Like in the collinear theory, all spins were 
initially restricted to point in the z-direction.
The wavefunctions and energies of the final result turned out to be identical 
with the ground state as computed in a separate collinear approach although the 
quantization axis of both atoms had rotated by $\pm$ 90 degrees during 
the iteration. This confirms the degeneracy of the electronic properties
with respect to the orientation of their spin. Furthermore we have been able
to verify the result of Car and coworkers\cite{car} concerning the 
noncollinear spin arrangement of Fe$_3$.

\begin{figure}[h,p,t]      
\unitlength1mm
\begin{picture}(180,200)
\put(-5.0,147.0){\parbox{75cm}{\psfig{figure=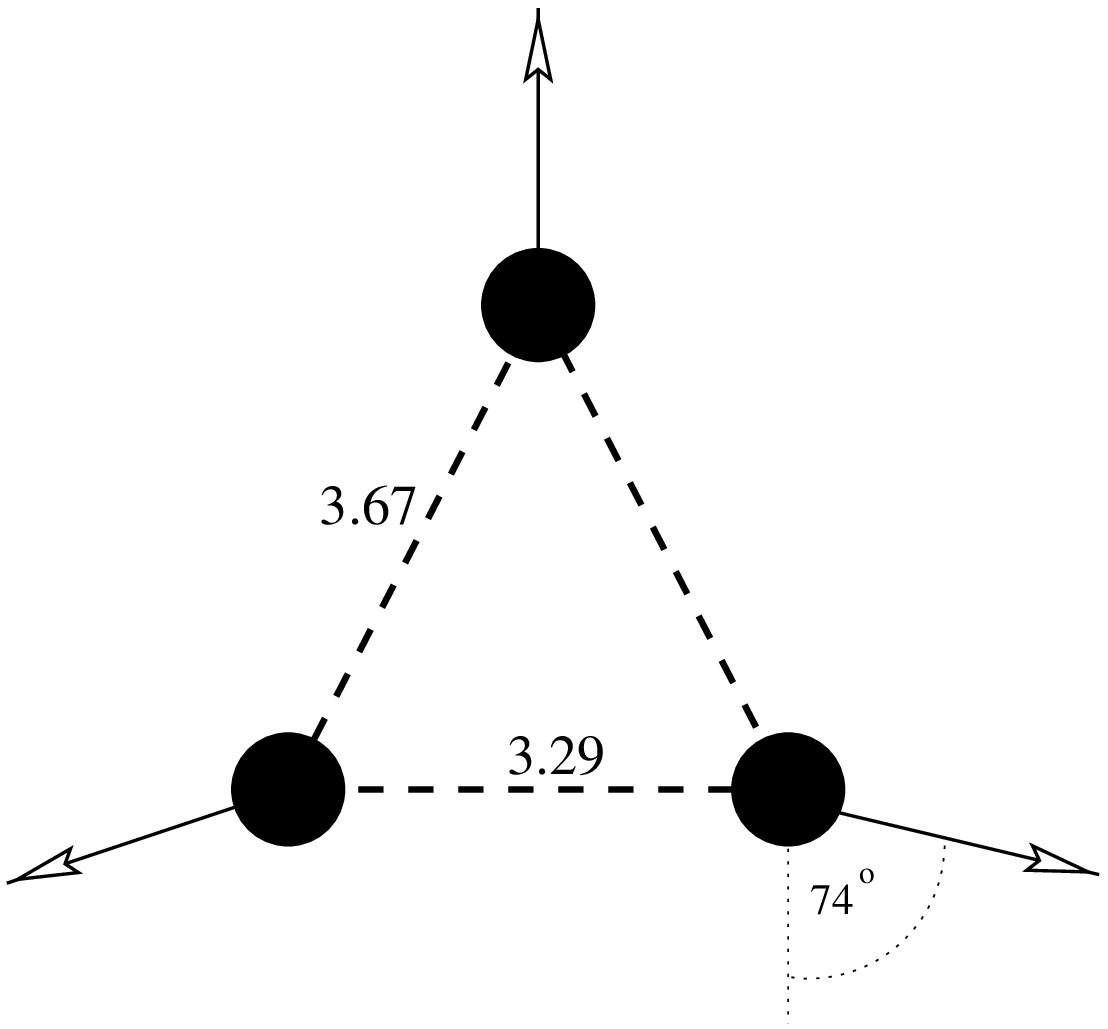,width=75mm}}}
\put(6.0,50.0){\parbox{5cm}{\psfig{figure=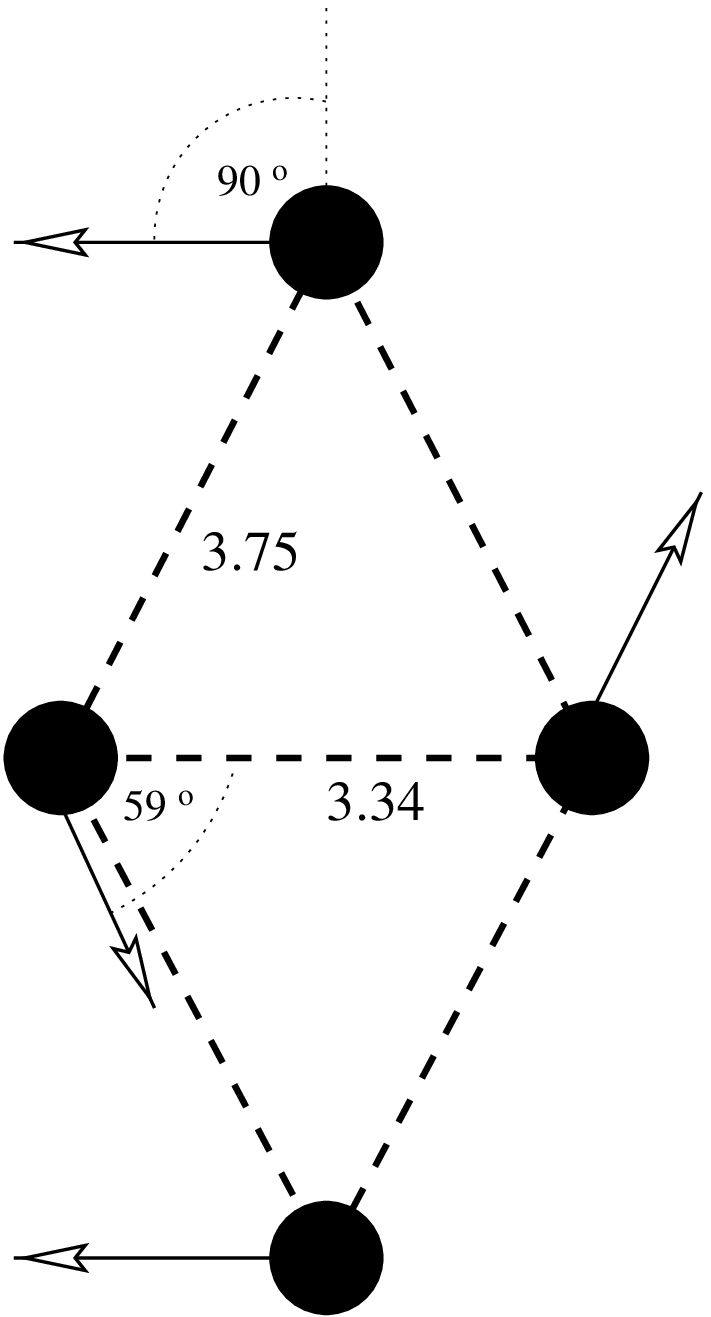,width=50mm}}}
\put(72.0,143.0){\parbox{104cm}{\psfig{figure=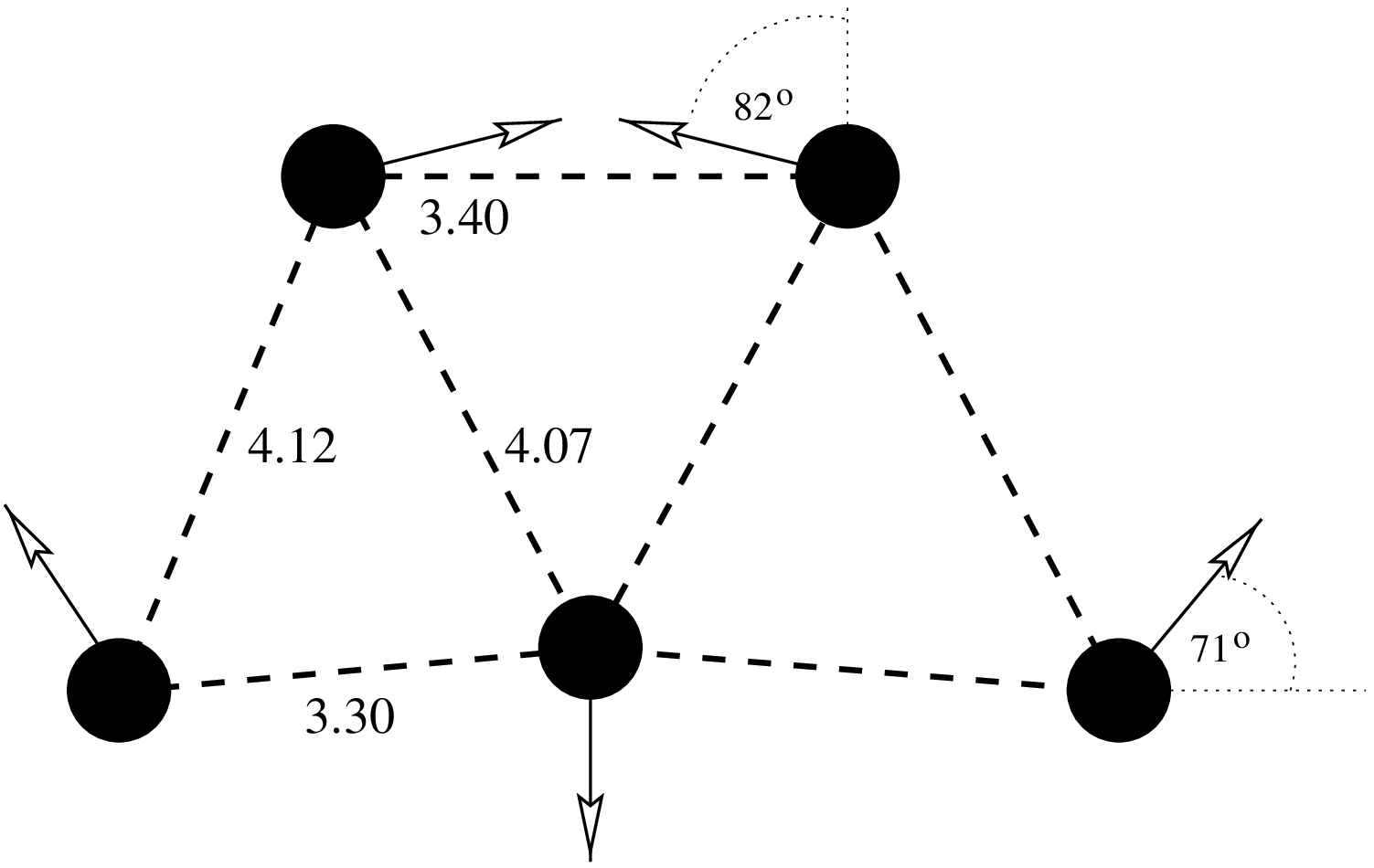,width=104mm}}}
\put(80.0,48.0){\parbox{90cm}{\psfig{figure=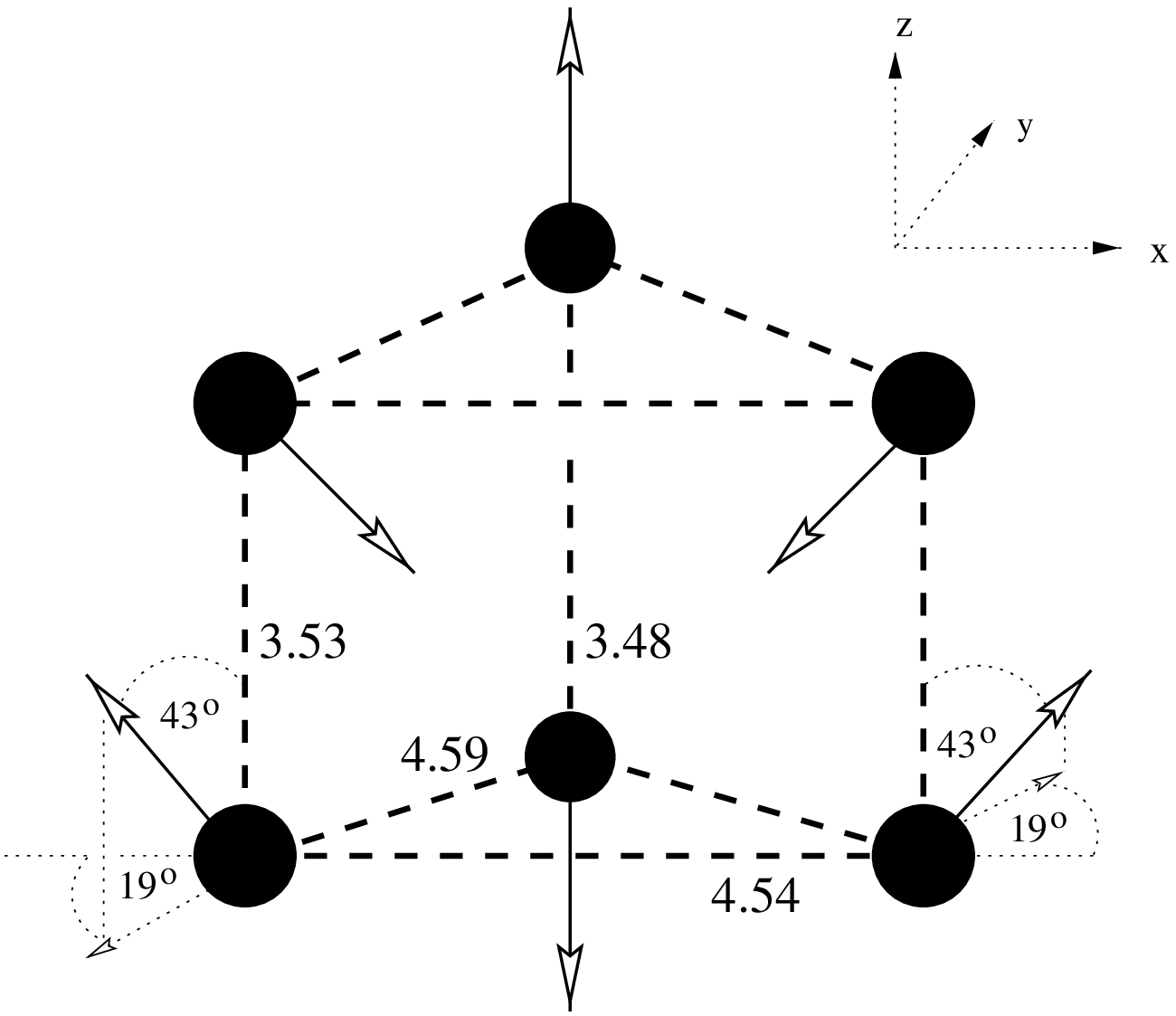,width=90mm}}}
\end{picture}
\begin{center}
\begin{minipage}{0.8\textwidth}
\vspace{0.8cm}
\caption{{\label{cr3-6}}
Geometric and magnetic structures for the energetically lowest 
noncollinear configurations of Cr$_{\rm N}$, $3 \le N 
\le 6$. The local magnetic moments including their angles with respect
to the x- and z-axis are indicated by arrows. The interatomic distances
are shown in atomic units.}
\end{minipage}
\end{center}
\end{figure}

\begin{figure}[h,p,t]
%\psdraft
\begin{center}
\begin{minipage}{1.0\textwidth}
\psfig{figure=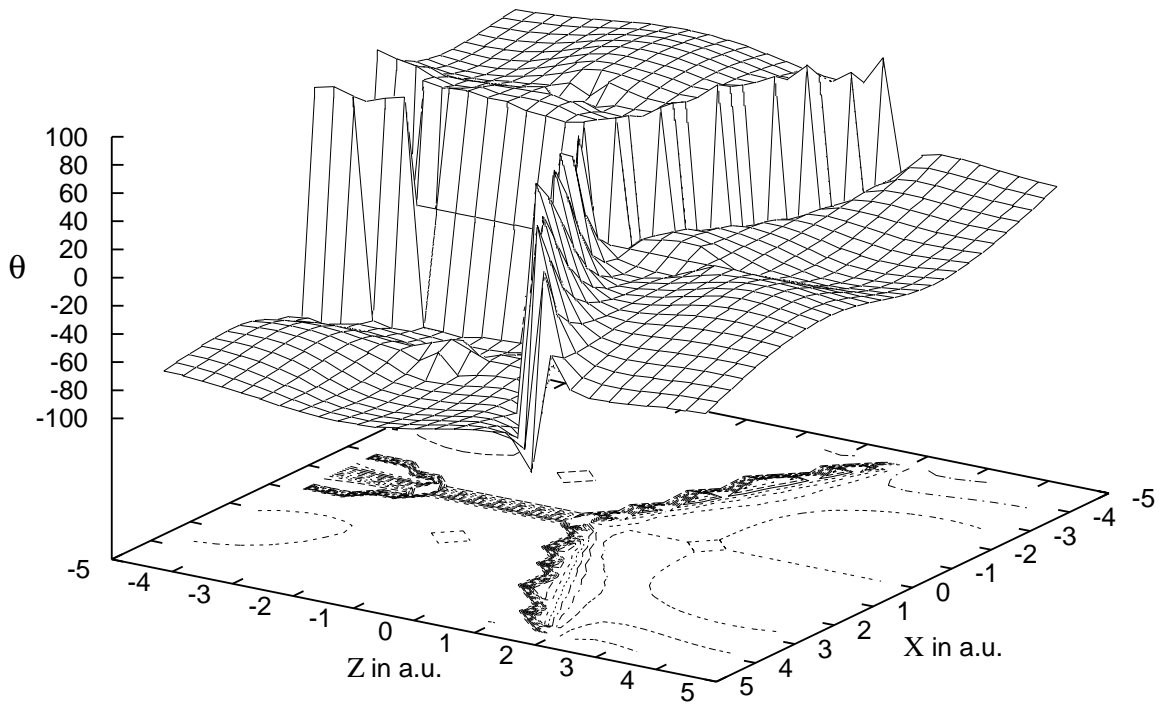,angle=-90,width=\textwidth}
%psfull
\end{minipage}
\vspace{4mm}
\begin{minipage}{0.85\textwidth}
\caption{{\label{3dplot}}
3D-plot of the polar rotation angle $\theta$ (in degrees) from 
eqn.(\ref{theta}) within the plane defined by the ionic coordinates of 
Cr$_{\rm 3}$ (see Fig.\ref{cr3-6}). We also show the corresponding contour 
lines with a step size of 10$^{\,o}$.}
\end{minipage}
\end{center}
\end{figure}

Our results for the energetically lowest noncollinear configurations of small 
chromium clusters Cr$_{\rm N}$ (3 $\le$ N $\le$ 6) are presented in 
Fig.\ref{cr3-6}. All structures except the one of Cr$_4$ represent ground 
states. The corresponding total magnetization is shown in Table~\ref{moments}.
Our geometric and magnetic structures are obtained by 
performing up to 50 full Monte-Carlo runs per cluster starting from arbitrary
ionic coordinates. A couple of thousand Kohn-Sham iterations are usually
necessary to completely relax the electronic degrees of freedom. This is 
because the numerical convergence with respect to the direction of the local
moments (governed by competing interatomic exchange interactions) is much 
slower than with respect to their magnitude which is determined by stronger 
intra-atomic interactions.\\
The principle effect which leads to noncollinear arrangements in chromium
clusters can best be demonstrated in Cr$_3$. Our calculation restricted to
collinear spins gives a trimer that basically consists of a dimer plus a 
loosely attached third atom, very similar to the result of Cheng and 
Wang~\cite{cheng}. It is obviously impossible for the atoms to couple 
antiferromagnetically with all their neighbors at a time so that a FE 
interaction between two atoms becomes inevitable -- a situation that we refer 
to in the following as frustration. The repulsive effect of the FE coupling at
smaller distances pushes one atom away and results in a large interatomic 
distance of 4.97 a.u..
If the noncollinear channel is accessible, however, the trimer is free to 
achieve the closest to AF-like coupling it can by rotating two local magnetic 
moments in the x-z plane (see Fig.\ref{cr3-6}). As the frustration is 
now reduced, the third atom comes much closer (d = 3.67 a.u.) and enables a 
buildup of molecular orbits which enhances the total binding inside the 
trimer significantly. The associated reduction of $|\vec{\mu}_{\rm at}|$ for 
the formerly isolated atom results in a decrease of M$_{\rm at}$ 
from $2 \, \mu_{\rm B}$ to $0.69 \,\mu_{\rm B}$. The 
energy difference with respect to the collinear ground state $\Delta 
{\rm E_{nc}}$ is 0.083 eV per atom (see Table~\ref{moments}) which amounts to
$7.8 \,\%$ of the binding energy of Cr$_3$. A look a the interatomic distances
and the tilted local moments suggests that the dimer lost its dominant role. 
However, the fact that the ground state is not an equilateral triangle with 
angles of 120$^o$ between the local magnetic moments (this would equal the 
best possible AF-like coupling) indicates that some trace of the strong dimer 
binding from the collinear calculation still persists. Unlike the situation
in Fe$_3$, the linear isomer of Cr$_3$ did not favor noncollinear spins.\\
The noncollinear spin structure of Cr$_3$ can be understood as a compromise 
between the energetically very favorable AF coupling and frustration. 
This can be seen easily with the very simple Hamiltonian $H = \beta 
\sum_{i < j}^3 \vec{\sigma}_i \cdot \vec{\sigma}_j$ for three spins on an 
equilateral triangle. Here $\beta$ is negative for AF coupling. In a collinear
restriction, the lowest energy state formed by a product wavefunction is 
$(\uparrow \uparrow \downarrow$) with an energy expectation $\langle H \rangle
= \beta$. Taking instead the state with $120^o$ angles between the spin 
directions gives a lower energy of $\langle H \rangle = 3/2 \, \beta$. 
For higher atomic spins (as they occur in our numerical computations) the
preference of the noncollinear configuration would of course become more 
pronounced due to the larger number of exchange interactions.\\
In Fig.\ref{3dplot} we show the rotation angle $\theta(\vec{r})$ in the x-z 
plane of the trimer including its contour lines. The azimuthal angle 
$\phi(\vec{r}$) is zero at any gridpoint due to the magnetic anisotropy
energy although this quantity is rather small. But it seems that even a 
difference of a few meV per atom is enough to keep the magnetization inside 
the plane of the trimer, an observation which agrees with the findings of 
ref.\cite{aniso}. The orientation of the magnetization density vector is 
remarkably uniform in the regions surrounding the atoms ($\theta \simeq \pm 
74^o, 0^o$ with a small oscillation of $\Delta\theta = \pm 10^o$) and varies 
significantly only right between them where the charge density is essentially 
zero. This is why a smaller integration radius only influences the magnitude 
of $\vec{\mu}_{\rm at}$ and not its orientation. The rapid change from 
$\theta=-90^o$ to $\theta = 90^o$ is related to a spin flip in these 
interatomic regions. We find an intra-atomic spin dispersion of around $6^o$ 
which partly comes from spin-orbit coupling. But the dispersion is 
also induced by the trend to an AF coupling with the neighbors. The same 
especially holds for the variation of the spin direction close to the domains 
of other atoms. The change of $\theta$ at the ionic positions 
is an artifact of the pseudopotential approximation. A detailed analysis 
further shows that these features are common to all investigated chromium 
clusters.\\
The collinear ground state of Cr$_4$ was found to have a rectangular structure 
(not shown) with bonding lengths of 3.35 a.u. and 4.62 a.u.. The double dimer 
structure of this geometry appears to be resistant to non-collinear effects 
as the distance of two atoms with parallel moments is very large (5.7 a.u.). 
But in the case of the lowest isomer, a rhombus, frustration becomes 
important again resulting in noncollinear spins (see Fig.\ref{cr3-6}). The 
argumentation follows the case of Cr$_3$, the triangles 
in Cr$_4$ are very similar and have only slightly larger bonding lengths. A 
higher total magnetization of M$_{\rm at} = 1.33 \, \mu_{\rm B}$ leads to an 
energy gain of 0.12 eV per atom with respect to the corresponding collinear 
state of the rhombus and halves the energy difference to the rectangular 
ground state. 
The isomer of Cr$_4$ is unique in that the energetically favored noncollinear 
configuration has a bigger total magnetization than its collinear counterpart. 
The magnitudes of $\vec{\mu}_{\rm at}$, however, were almost identical in 
both cases. It is also noteworthy that the rhombus structure provides angles 
between the local moments of almost exactly $120^o$ although the bondings are 
not equivalent. This seems to be related with the larger moments of the
furthermost atoms ($\Delta|\vec{\mu}_{\rm at}| = 0.6 \, \mu_{\rm B}$).\\ 
The properties of the noncollinear ground state of Cr$_5$ can be understood  
if one considers the pentamer as consisting of three triangles. Although the
bonding lengths are bigger, each of these triangles exhibits the same 
basic features as Cr$_3$ and the local moments are again arranged in such a 
way that the best possible AF-like coupling is achieved. We also see that
the magnitude of $\vec{\mu}_{\rm at}$ decreases with increasing coordination 
number. Note that the reflection symmetry of the geometry is the same as the
symmetry of the magnetization. The gain in binding energy with respect to the 
best collinear state is 0.054 eV per atom ($=3.7\,\%$ of the binding energy)
which is less than for the trimer. But the total magnetization is again 
clearly reduced from 1.03 to $0.53 \, \mu_{\rm B}$ per atom. The lowest isomer
is a bipyramid with a noncollinear spin structure as well and M$_{\rm at} 
= 0.79 \, \mu_{\rm B}$. Its energy difference of 0.6 eV to the ground state is 
0.25 eV higher than in the collinear case. Cr$_5$ is the only cluster we found
where the noncollinear and collinear geometries differ by more than 
just a variation of the bonding lengths. Our lowest collinear state looks 
similar to the geometry of Cheng and Wang and has a 2v-symmetry. However, its 
total magnetization of $1.05 \, \mu_{\rm B}$ is somewhat higher than their 
result of $0.93 \, \mu_{\rm B}$.\\ 
The shape of Cr$_6$, on the other hand, resembles the collinear one very 
closely. Cr$_6$ is the smallest cluster that displays a fully three 
dimensional geometric and magnetic structure in the ground state. It consists 
of three dimers distributed over two triangles in which frustration sets in. 
Each of the $\vec{\mu}_{\rm at}$ is exactly antiparallel to the moment of its 
partner atom in the dimer. The azimuthal angles are $\phi = \pm
19^o$ for the atoms in the foreground and $\phi = 0^o$ for the ones in the 
background which are slightly closer. The bonding lengths in the triangles 
are about $6 \, \%$ shorter than in the collinear case and the dimer distances 
are somewhat bigger. It is important to note that in spite of a vanishing total
magnetization and although the dimer seems to recover a certain influence, 
tilted spins are still energetically favored. However, the gain 
of $\Delta {\rm E_{nc}} = 0.022$ eV per atom only accounts for $1.4 \,\%$ of 
the binding energy of Cr$_6$. This indicates that the impact of noncollinear 
effects on the energetics becomes gradually smaller with rising cluster size. 
Further calculations for Cr$_7$ and Cr$_9$ confirm this trend and predict that 
the noncollinear geometries essentially equal the collinear ones for N$_{\rm 
at} \ge 6$. On the other hand our results so far show that noncollinear spin 
configurations considerably reduce the total magnetization of the ground states
although it might not be reflected too much in the binding energy. This can 
be understood in terms of the competing interatomic exchange interactions 
which involve only small energy changes during the rotation of local moments 
(see above).

\begin{figure}[h,p,t]
%\psdraft
\begin{center}
\begin{minipage}{1.0\textwidth}
\psfig{figure=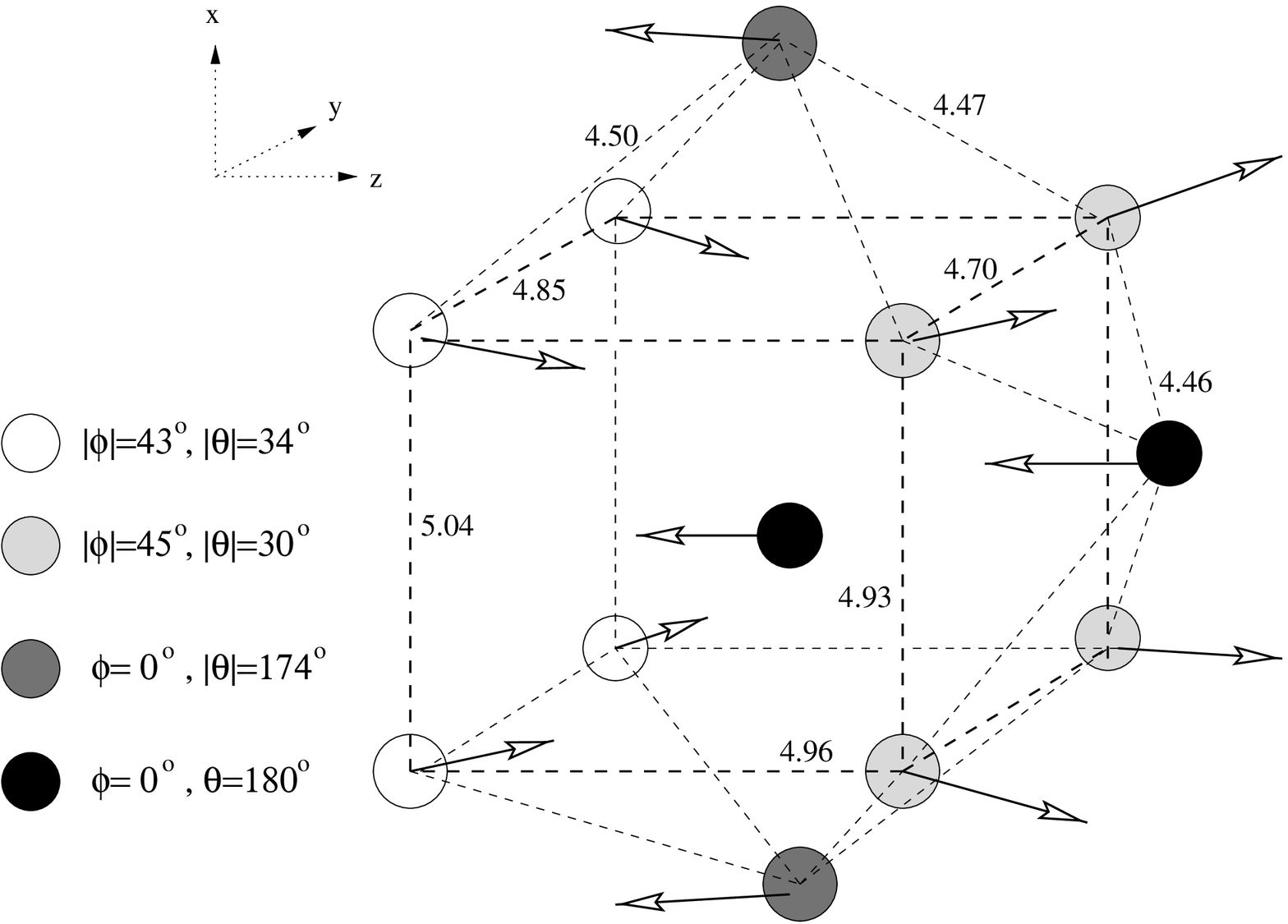,width=\textwidth}
%psfull
\vspace{2cm}
\end{minipage}
\begin{minipage}{0.8\textwidth}
\caption{{\label{cr12}}
Geometric and magnetic structure of the noncollinear ground state 
of Cr$_{\rm 12}$. The notation follows the one from Fig.\ref{cr3-6}.
The magnetic moments of the lighter-colored atoms show a dispersion
in the positive z-direction whereas the darker atoms almost exactly 
point in the negative z-direction. We also display the absolute values 
of the azimuthal and polar angles of $\vec{\mu}_{\rm at}$ for all 4 
groups of atoms.} 
\end{minipage}
\end{center}
\end{figure}

We shall finally discuss how noncollinear effects might provide a way to 
ameliorate the disagreement between the measured total magnetic moments 
of Cr$_{\rm N}$ (N $\ge$ 9) by Bloomfield et al.\cite{bloom} and previous 
theoretical results\cite{cheng,muprev}. The Stern-Gerlach experiment extracted
an upper bound for M$_{\rm at}$,  M$_{\rm at} < 0.77 \, \mu_{\rm B}$, assuming
a superparamagnetic behavior of the chromium clusters. Earlier calculations, 
however, report values for the magnetization of some clusters that are  
much higher\cite{muprev}. The values from Cheng and Wang for Cr$_{12}$ 
(M$_{\rm at} = 1.67 \, \mu_{\rm B}$) and Cr$_{13}$ (M$_{\rm at} = 1.06 \,
\mu_{\rm B}$) also exceed the experimental limit. The trend from smaller 
clusters gives one the hope that noncollinear effects might reduce the 
difference.
However, an unconstrained simultaneous calculation of electronic and ionic 
degrees of freedom surpasses our computational resources for clusters as large
as Cr$_{12}$. Instead, we start the optimization procedure from the
geometries of Cheng and Wang. This is a reasonable approach in the light
of the very good agreement with our collinear structures and the small changes
in geometry that are induced by noncollinear spins. Our final magnetic and 
geometric configuration of Cr$_{12}$ is shown in Fig.\ref{cr12}. The free 
relaxation leads to a shortening of the bonding lengths between the corner 
atoms of about $5 \,\%$ and slightly bigger distances of the capping atoms but
the bulk-like bcc-structure of the collinear geometry clearly persists.
The z-components of the local magnetic moments vary on alternating x-y planes,
but only the moments of the corner atoms have significant x- and y-components.
All the spins of the lightest-colored corner atoms point inside the cube 
towards the central atom, whereas the spins of the slightly darker corner 
atoms point outside and away from the next atom. This dispersion is related to
some frustration of the corner atoms in connection with the preferred AF 
coupling to the moments of their nearest neighbors (darker atoms). A magnetic 
arrangement like that can be seen as a precursor to the bulk behavior in form 
of an antiparallel order between neighboring atomic layers. The tilted spins 
of the corner atoms ($\phi = \pm 43^o, \theta=\pm 34^o \, \mbox{and}\, 
\phi=\pm 45^o, \theta=\pm 30^o$) have pairwise opposite x- and y-components so
that the sum of their net-moments in positive z-direction is reduced. The 
moments of the darker atoms, however, show almost no dispersion in negative 
z-direction. All this results in a much smaller total magnetization of 
M$_{\rm at} = 0.81 \, \mu_{\rm B}$ which is now very close to the experimental
limit. The remarkable reduction of $\Delta {\rm M_{at}} = -0.86 \, 
\mu_{\rm B}$ is associated with $\Delta {\rm E_{nc}}$ = 0.011 eV per atom 
which equals only $0.5\,\%$ of the binding energy of Cr$_{12}$. A very similar
situation leads to M$_{\rm at} = 0.59 \, \mu_{\rm B}$ in the case of 
Cr$_{13}$, a value that is even below the experimental limit. The noncollinear
gain amounts to $\Delta {\rm E_{nc}}$ = 0.008 eV per atom. The Cr$_{13}$ result
is preliminary and might slightly change ($\Delta {\rm M_{at}} \le \pm \, 
0.05 \, \mu_{\rm B}$) because the electronic degrees of freedom were not 
completely relaxed. But we can still conclude that the total magnetization of 
Cr$_{12}$ and Cr$_{13}$ as obtained with the general LSDA represents a 
considerable improvement with respect to the experiment.

\section{Conclusions}
We present the first study of noncollinear effects in antiferromagnetically
coupled clusters by applying the general, rotationally invariant
LSDA for the electronic degrees of freedom. Their interaction with the ions is 
described in terms of a relativistic, non-local pseudopotential which has been
thoroughly tested. The magnetic and geometric structures are obtained by 
employing a simulated annealing technique for the ionic optimization 
together with interlaced Kohn-Sham iterations which update the wavefunctions 
simultaneously.\\
Our collinear configurations, which we use in order to extract the effect of
noncollinear spins, agree very well with the results of Cheng and 
Wang\cite{cheng}. We find that all investigated chromium clusters show a 
pronounced trend to noncollinear spin configurations. This is caused by a 
subtle interplay between the preferred magnetic order and frustration, 
a situation which can in principle occur in all clusters of elements that 
favor antiferromagnetic spins. Therefore our considerations appear 
to be of a more general nature although the special properties of chromium 
indicate that noncollinear effects could be less dramatic in other 
transition metals. The results for N$_{\rm at} \le 13$ show that the 
influence of noncollinearity on various observables becomes gradually smaller 
with rising cluster size. Induced changes of the cluster geometry are  
generally restricted to an alteration of the bonding lengths, with the 
exception of Cr$_5$. However, we observe a universal reduction of the total 
magnetization which is significant even in those clusters for which a 
variation of the magnetic distribution is not clearly reflected in the 
binding energy any more (Cr$_{12}$ and Cr$_{13}$). This effect is related to 
the small energetic changes which occur during the rotation of local moments, 
an aspect that makes a proper convergence of the Kohn-Sham iteration very time 
consuming. Furthermore, our findings show that tilted spins due to 
frustration can even be favored when the total magnetization vanishes (Cr$_6$)
or the corresponding ground state is collinear like in Cr$_4$. 
The free variation of the spin quantization axis finally leads to a better 
agreement with the experiment concerning the total magnetization of Cr$_{12}$ 
and Cr$_{13}$. It can be concluded that noncollinear effects appear to be an
important ingredient for a deeper understanding of the subtle magnetic 
properties in transition metal clusters.\\

\noindent {\bf Acknowledgments:}
One of the authors (C.K.) has been supported by the DAAD (German Academic
Exchange Service), Grant No. D/98/14581. We also thank Ana Proykova, Sanjay 
Reddy, Paul-Gerhard Reinhard, Louis Bloomfield and Lai-Sheng Wang 
for many useful and encouraging discussions.

%\newpage
%\mbox{}
%\newpage

%\newpage
%\mbox{}
\newpage

\mbox{}

\begin{table}[h,t,p]
\begin{center}
\begin{minipage}{0.7\textwidth}
\begin{tabular}  {lllll} 
N$_{\rm at}$ & Type & M$_{\rm at}$ & $\Delta {\rm E_{nc}}$ & M$_{\rm at}$ 
(ref.\cite{cheng}) \\ \hline
2      & collinear    &  0.0 (\mbox{--}) & 0.0 (\mbox{--}) & 0.0 \\
3      & noncollinear &  0.69            & 0.083    & 2.0  \\
4      & collinear    &  0.0 (1.33)      & 0.0 (0.12) & 0.0 \\ 
5      & noncollinear &  0.53            & 0.054    & 0.93 \\
6      & noncollinear &  0.0             & 0.022    & 0.33 \\
7      & noncollinear &  0.13            & 0.019    & 0.29 \\
9      & noncollinear &  0.09            & 0.015    & 0.22 \\
12     & noncollinear &  0.81            & 0.011    & 1.67 \\
13$^{\ast}$ & noncollinear &  0.59            & 0.008    & 1.06 \\
\end{tabular}
\vspace{4mm}
\caption{{\label{moments}}
Total magnetization per atom M$_{\rm at}$ (in units $\mu_{\rm B}$) for the 
ground states of Cr$_{\rm N}$ and gain in binding energy $\Delta {\rm E_{nc}}$
(in units eV/atom) with respect to their collinear counterparts. 
In case of a collinear ground state the result for the energetically lowest 
noncollinear isomer is given in parentheses. The last column shows the 
corresponding magnetization from the collinear calculation of Cheng and 
Wang\cite{cheng}.
The star at N$_{\rm at}=13$ indicates that the structure was not completely 
relaxed.}
\end{minipage}
\end{center}
\end{table}

%\newpage 
%\mbox{}
\newpage

\mbox{}
\vspace{2cm}

\begin{thebibliography}{100}

\bibitem{exp1}
D.C. Douglass, A.J. Cox, J.P. Bucher, L.A. Bloomfield:
Phys.Rev. B {\bf 47}, 12874 (1993); 
S.E. Ampsel, J.W. Emmert, L.A.  Bloomfield: Phys.Rev. Lett. {\bf 76}, 1441 
(1996)

\bibitem{exp2}
I.M.L. Billas, J.A. Becker, W.A. de Heer: Z.Phys. D {\bf 24}, 325 (1993);
I.M.L. Billas, A. Ch\^{a}telain, W.A. de Heer: Science {\bf 265}, 1682 (1994)

\bibitem{exp3}
G. Gantef\"or, W. Eberhardt: Phys.Rev.Lett. {\bf 76}, 4975 (1996)

\bibitem{exp4}
H. Wu, S.R. Desai, L. Wang: Phys.Rev.Lett. {\bf 77}, 2436 (1996)

\bibitem{theo1}
B.V. Reddy, S.N. Khanna, B.I. Dunlap: Phys.Rev.Lett. {\bf 70}, 3323 (1993);
M. Menon, J, Connolly, N.N. Lathiotakis, A.N. Andriotis: Phys.Rev. B 
{\bf 50}, 8903 (1994) 
  
\bibitem{theo2}
G.F. Bertsch, N. Onishi, K. Yabana: Z.Phys. D {\bf 34}, 213 (1995);
V. Visuthikraisee, G.F. Bertsch: Phys.Rev. A {\bf 54}, 5104 (1996)

\bibitem{theo3}
J. Guevara, F. Parisi, A.M. Llois, M. Weissmann: Phys.Rev. B {\bf 55}, 13283
(1997);
Q. Sun, G. Wang, J.Z. Yu, Z.Q. Li, J.T. Weng: J.Phys. I France {\bf 7}, 1233 
(1997) 

\bibitem{theo4}
A.N. Andriotis, N.N. Lathiotakis, M. Menon: Europhys.Lett. {\bf 36}, 37 
(1996)

\bibitem{rhodium}
A.J. Cox, J.G. Louderbeck, S.E. Apsel, L.A.  Bloomfield: Phys.Rev. B {\bf 49}, 
12295 (1994); 
P. Villase\~{n}or-Gonz\'{a}lez, J. Dorantes-D\'{a}villa, H. Dreyss\'{e},
G.M. Pastor: Phys.Rev. B {\bf 55}, 15084 (1997)

\bibitem{nickel}
S.E. Apsel, J.W. Demmert, J. Deng, L.A. Bloomfield: Phys.Rev.Lett. {\bf 76},
1441 (1996); 
I.M.L. Billas, A. Ch\^{a}telain, W.A. de Heer: J.Magn.Mater. {\bf 168}, 64 
(1997); 
F. Aguilera-Granja, S. Bouarab, M.J. L\'{o}pez, A. Vega, J.M. 
Montejano-Carrizales, M.P. I\~{n}iguez, J.A. Alonso: Phys.Rev. B {\bf 57},
12469 (1998)
 
\bibitem{aniso}
G.M. Pastor, J. Dorantes-D\'{a}villa, S. Pick, H. Dreyss\'{e}:
Phys.Rev.Lett. {\bf 75}, 326 (1995); 
J. Dorantes-D\'{a}villa, G.M. Pastor: Phys.Rev.Lett. {\bf 81}, 208 (1998) 

\bibitem{temp}
P.J. Jensen, K.H. Bennemann: Z.Phys. D {\bf 35}, 273 (1995); 
J. Zhao, X. Chen, Q. Sun, F. Liu, G. Wang: Europhys.Lett. {\bf 32}, 113 (1995);
P. Borrmann, B. Diekmann, E.R. Hilf, D. Tom\'{a}nek: Surf. Rev. \& Lett. 
{\bf 3}, 463 (1996);
M. Respaud, J.M. Broto, H. Rakato, A.R. Fert, L. Thomas, B. Barbara,
M. Verelst, E. Snoeck, P. Lecante, A. Mosset, J. Osuna, T.O Ely,
C. Amiens, B. Chaudret: Phys.Rev. B {\bf 57}, 2925 (1998)

\bibitem{car}
T. Oda, A. Pasquarello, R. Car: Phys.Rev.Lett. {\bf 80}, 3622 (1998)

\bibitem{bulkfe}
M. Uhl, L.M. Sandrtskii, J. K\"ubler: Phys.Rev. B {\bf 50}, 291 (1994)

\bibitem{disord}
M Liebs, K. Hummler, M. F\"ahnle: Phys.Rev. B {\bf 51}, 8664 (1995);
R. Lorenz, J. Hafner, S.S. Jaswal, D.J. Sellmyer: Phys.Rev.Lett. {\bf 74}, 
3688 (1995)

\bibitem{nonfer}
R. Lorenz, J. Hafner: Phys.Rev. B {\bf 58}, 5197 (1998)

\bibitem{multi1}
A. Schreyer, C.F. Majkrzak, T. Zeidler, T. Schmitte, P. B\"odecker,
K. Theis-Br\"ohl, A. Abromeit, J.A. Dura, T. Wanabe: Phys.Rev.Lett. {\bf 79}, 
4914 (1997);
R.S Fishman: Phys.Rev.Lett. {\bf 81}, 4979 (1998)

\bibitem{multi2}
S. Demuyck, J. Meerschaut, J. Dekoster, B. Swimmen, R. Moons,
A. Vantomme, S. Cottenier, M. Rots: Phys.Rev.Lett. {\bf 81}, 2562 (1998)

\bibitem{embedd}
P. Alvarado, J. Dorantes-D\'{a}villa, G.M. Pastor: Phys.Rev. B {\bf 58}, 12216
(1998)
  
\bibitem{mann}
E. Viitala, J. Merikoski, M. Manninen, J. Timonen: Phys.Rev. B {\bf 55}, 11541
(1997)

\bibitem{dimerexp}
S.M. Casey, D.G. Leopold: J.Chem.Phys. {\bf 97}, 816 (1993)
and references therein

\bibitem{dimer}
B. Delley, A.K. Freemann, D.E. Ellis: Phys.Rev.Let. {\bf 50}, 488 (1983);
J. Bernholc, N.A.W Holtzwarth: Phys.Rev. Lett. {\bf 50}, 1451 (1983);
L. Andersson: Chem.Phys.Lett. {\bf 237}, 212 (1995)

\bibitem{cheng}
H. Cheng, L. Wang: Phys.Rev.Lett. {\bf 77}, 51 (1996)

\bibitem{bloom}
D.C. Douglass, J.P. Bucher, L.A. Bloomfield: Phys.Rev. B {\bf 45}, 6341 (1992)

\bibitem{muprev}
G.M. Pastor, J. Dorantes-D\'{a}villa, K.H. Bennemann: Phys.Rev. B 
{\bf 40},7642 (1989);
K. Lee, J. Callaway: Phys.Rev. B {\bf 49}, 13906 (1994)

\bibitem{hedin}
U. Barth, L. Hedin: J.Phys. C, Solid State Phys. {\bf 5}, 1629 (1972)

\bibitem{perdew}
J.P. Perdew, Y. Wang: Phys.Rev. B {\bf 45}, 13244 (1992)

\bibitem{kueb}
J. K\"ubler, K.-H. H\"ock, J. Sticht, A.R. Williams: J.Phys F, Met.Phys.
{\bf 18}, 469 (1988)

\bibitem{fukutome}
H. Fukutome: Int.J.Quant.Chem. {\bf 20}, 955 (1981)

\bibitem{goed}
C. Hartwigsen, S. Goedecker, and J. Hutter: Phys.Rev. B {\bf 58}, 3641 (1998)

\bibitem{asa}
L. Nordstr\"om. D.J. Singh: Phys.Rev.Lett. {\bf 76}, 4420 (1996)

\bibitem{diss}
C. Kohl: Ph.D. thesis, Erlangen-Germany, 1997

\bibitem{damp}
V. Blum, G. Lauritsch, P.-G. Reinhard, J.A. Maruhn: J.Comp.Phys. {\bf 100},
364 (1992)

\bibitem{falr}
G. Lauritsch, P.-G. Reinhard: Int.J.Mod.Phys. C {\bf 5}, 65 (1994)

\bibitem{numasp}
C. Kohl, B. Fischer, P.-G. Reinhard: Phys.Rev. B. {\bf 56}, 11149 (1997);
C. Kohl, F. Calvayrac, P.-G. Reinhard, E. Suraud: Surf. Science {\bf 405}, 74
(1998)

\end{thebibliography}
\end{document}